\documentstyle[twocolumn,aps,epsfig]{revtex}
\begin{document}
\draft
\title{
Singularities and Pseudogaps in the
Density of  States of Peierls Chains}

\author{Lorenz Bartosch and Peter Kopietz} 
%\author{Peter Kopietz} 
%\author{PK's preliminary draft}
%
\address{
Institut f\"{u}r Theoretische Physik, Universit\"{a}t G\"{o}ttingen,
Bunsenstrasse 9, D-37073 G\"{o}ttingen, Germany}
\date{October 23, 1998}
\maketitle
\begin{abstract}
We develop a non-perturbative method to 
calculate the density of states
(DOS) $\rho ( \omega )$ of the fluctuating gap model  
describing the low-energy physics of electrons on a disordered
Peierls chain. 
For  a real order parameter field
we calculate $\rho (  0 )$ (i.e.
the DOS at the Fermi energy) {\it{exactly}}
as a functional of the disorder for 
a chain of finite length $L$.
Averaging $\rho ( 0)$ with respect to a Gaussian
probability distribution of the fluctuating
Peierls order parameter,
we show that for $L \rightarrow \infty$ the
average
$\langle \rho ( 0 ) \rangle$  
diverges for any finite value of the correlation length
above the Peierls transition.
Pseudogap behavior emerges only if the
Peierls order parameter is finite and sufficiently large.
\end{abstract}
\pacs{PACS numbers: 71.23.-k, 02.50.Ey, 71.10.Pm}
\narrowtext
%
%    I N T R O D U C T I O N
%
%\section{Introduction}
%\setcounter{section}{1}
%\label{sec:intro}
%

At low temperatures 
many quasi one-dimensional conductors 
become unstable and develop  long-range
charge-density-wave order, i.e. undergo a Peierls-transition\cite{Gruener94}.
Within a mean field picture 
a finite value $\Delta_0$ of the Peierls
order parameter leads for frequencies 
$ | \omega | < |\Delta_0|$ 
to a gap
in the electronic density of states (DOS)
$ \rho ( \omega  )$\cite{footnote1}. 
To achieve a better understanding of
the effect of fluctuations on the Peierls
transition, Lee, Rice, and Anderson\cite{Lee73}
introduced the so-called fluctuating gap model
(FGM).
In this model
the fluctuating part $\tilde{\Delta}  ( x ) = \Delta ( x ) - \Delta_0$
of the order parameter is approximated by a Gaussian stochastic process
with covariance
 \begin{equation}
  \langle   \tilde{\Delta} (x ) 
  \tilde{\Delta} ( x^{\prime} ) \rangle
  \equiv 
  K ( x , x^{\prime} ) 
  = K_0
  e^{ - | x - x^{\prime} | / \xi } 
  \; .
  \label{eq:cov}
  \end{equation}
Here $\langle \ldots \rangle$
denotes averaging over the 
probability distribution of 
$\Delta (x )$, $K_0$ is a positive constant, and 
$\xi$ is the order parameter
correlation length.
We assume that the
field $\Delta (x)$ is real, 
corresponding to a charge-density wave that is commensurate
with the lattice.

Twenty years ago 
Sadovskii\cite{Sadovskii79}
found an apparently exact algorithm   
to calculate the average DOS of the FGM.
His calculations showed  that for temperatures
above the Peierls transition, in a regime where 
$\xi$ is large but finite, 
the average DOS exhibits a 
substantial suppression in the vicinity of the
Fermi energy, a so-called pseudogap.
The algorithm constructed
by Sadovskii has also been applied in a different context to
explain the weak pseudogap behavior in the
underdoped cuprates\cite{Schmalian98}. 
However, recently it has been pointed 
out\cite{Tchernyshydov98,Kuchinskii98}
that Sadovskii's algorithm contains a subtle flaw
and hence does  not produce the exact
DOS of the FGM. It is therefore important 
to compare this algorithm with
limiting cases where $\langle \rho ( \omega ) \rangle$
can be calculated without any approximation.

Besides the limit $\xi \rightarrow \infty$
where Sadovskii's algorithm is indeed 
exact\cite{Tchernyshydov98,Kuchinskii98},
there exists  another non-trivial limit where
the exact $\langle \rho ( \omega ) \rangle$ is known:
if in Eq.(\ref{eq:cov}) we let
$\xi \rightarrow 0$, $K_0 \rightarrow \infty$, with 
$K_0  \xi \rightarrow D = {\rm const}$, the
right-hand side of Eq.(\ref{eq:cov})
reduces to $2 D \delta ( x - x^{\prime})$. 
As shown by Ovchinnikov and
Erikhman (OE)\cite{Ovchinnikov77},
in the limit $L \rightarrow \infty$
(where $L$ is the length of the chain)
the exact $\langle \rho ( \omega ) \rangle$
can then be obtained
from the stationary solution of a Fokker-Planck 
equation\cite{Lifshits88}.
For small $\omega$ and $\Delta_0 = 0$ one finds\cite{Ovchinnikov77}
$\langle \rho ( \omega ) \rangle \propto
| \omega   \ln^3  | \omega |  |^{-1}$.
Singularities of this type at the band
center of a random Hamiltonian
have been discovered
by Dyson\cite{Dyson53}, 
and have recently also been found
in one-dimensional spin-gap systems\cite{Fabrizio97,McKenzie96}.
It is important to note that in the FGM the
singularity is a consequence of
the charge conjugation symmetry of the  
underlying Dirac Hamiltonian, 
and is {\it{not}} related to concrete probability properties
%of $\Delta (x)$\cite{Lifshits88,Mostovoy98,Eggarter78}. 
of $\Delta (x)$\cite{Lifshits88,Mostovoy98}. 
In particular, the singularity 
is {\it{not}} an artefact of the exactly solvable limit 
$\xi \rightarrow 0$ considered by OE\cite{Ovchinnikov77}. 
It is therefore reasonable to expect that
for any $\xi < \infty$ the average DOS of the FGM 
exhibits a singularity at $\omega = 0$.
This general argument is in disagreement
with Ref.\cite{Sadovskii79},
where for large but finite $\xi$ 
a pseudogap (and hence no singularity) has been obtained.
In this work we shall 
resolve this contradiction by calculating
the average DOS at the Fermi energy 
(i.e. $\langle \rho ( \omega = 0) \rangle$)
{\it{exactly for arbitrary $\xi$.}}

%In this work we present an {\it{exact}} calculation
%of the average DOS of the FGM at the Fermi energy
%{\it{for arbitrary $\xi$}}. 
%In the limit $L \rightarrow \infty$
%we find that
%$\langle \rho ( 0 ) \rangle$  diverges 
%as long as $ | \Delta_0 |$ is smaller than a
%finite critical value. 
%
%that indeed the DOS of the FGM with real
%$\Delta$ exhibits a Dyson singularity (and hence
%no pseudogap) for arbitrary $\xi < \infty$.
%so that one should expect that the average DOS
%$\langle \rho ( \omega = 0 ) \rangle$
%of the FGM exhibits a Dyson singularity
%for any finite value of the correlation length $\xi$.
%Obviously this is in disagreement with
%the results of Sadovskii\cite{Sadovskii79},
%who found that for sufficiently large $\xi$
%the average DOS of the FGM exhibits a pseudogap at the
%Fermi energy. 

The local DOS $\rho ( x , \omega )$ of the FGM
for a given realization of the disorder can be written 
as\label{eq:footnote1}
 \begin{equation}
 \rho ( x , \omega ) = - {\pi}^{-1}
 {\rm Im} {\rm Tr} [ \sigma_3
 {\cal{G}} ( x , x , \omega + i0 ) ]
 \; ,
 \label{eq:localdos}
 \end{equation}
where the $2 \times 2$ matrix Green's function
${\cal{G}}$ 
satisfies
 \begin{equation}
 \left[ i \partial_x + \omega \sigma_{3}
 - i \Delta ( x ) \sigma_{2}   \right]
 {\cal{G}} ( x , x^{\prime} , \omega )
 = \delta ( x - x^{\prime}) \sigma_0
 \label{eq:matrix}
 \; .
 \end{equation}
Here $\sigma_{i}$ are the usual Pauli matrices
and 
$\sigma_0$ is the $2 \times 2$ unit matrix. Note that in
Eq.(\ref{eq:localdos}) we have
factored out a Pauli matrix $\sigma_{3}$, so that the 
differential operator $i \partial_x$ 
in Eq.(\ref{eq:matrix}) is proportional
to the unit matrix.
To solve Eq.(\ref{eq:matrix}), we try the ansatz
(suppressing for simplicity the frequency label)
 \begin{equation}
 {\cal{G}} ( x , x^{\prime}  ) = U ( x )
 {\cal{G}}_1 ( x , x^{\prime}  )  U^{-1} ( x^{\prime} )
 \; ,
 \label{eq:ansatz}
 \end{equation}
where $U ( x )$ is an invertible $2 \times 2 $ matrix.
Eq.(\ref{eq:ansatz})
resembles the transformation law 
of the {\em comparator} in non-Abelian
gauge theories\cite{Peskin95}.
In fact,
Eq.(\ref{eq:ansatz}) can be viewed as a
gauge transformation which
generalizes the Schwinger ansatz\cite{Schwinger62}
to the non-Abelian case.
It is easy to show that the solution of
Eq.(\ref{eq:matrix}) can indeed be written in the form
(\ref{eq:ansatz}) provided
${\cal{G}}_1$ and $U$ satisfy
 \begin{eqnarray}
 \left[ i \partial_x + \omega \sigma_{3}
    \right]
 {\cal{G}}_1 ( x , x^{\prime}  )
 & = & \delta ( x - x^{\prime}) \sigma_0
 \label{eq:matrix1}
 \; ,
 \\
 & & \hspace{-39mm} i \partial_x U  (x )  =  \omega
 [ U ( x )  \sigma_3  - \sigma_3 U (x)] + 
 i \Delta ( x ) \sigma_{2}   U ( x )
 \; .
 \label{eq:Udif}
 \end{eqnarray}
%where $[ \; , \; ]$ denotes the commutator.
%For the derivation of Eqs.(\ref{eq:matrix1},\ref{eq:Udif})
%it is crucial that the
%differential operator $i \partial_x$ 
%in Eq.(\ref{eq:matrix}) is proportional
%to the unit matrix. This is the reason
%why in Eq.(\ref{eq:localdos}) we have factored out
%the Pauli matrix $\sigma_3$.
Eq.(\ref{eq:matrix1}) defines the Green's function of
free fermions, and can
be solved trivially via Fourier transformation.
The difficult part of the calculation 
is the solution of the matrix equation (\ref{eq:Udif}).
We parameterize $U ( x)$ as follows,
 \begin{equation}
 U ( x ) = 
 e^{i \Phi_{+} ( x ) \sigma_{-} }
 e^{i \Phi_{-} ( x ) \sigma_{+} }
 e^{i \Phi_{3} ( x ) \sigma_{3} }
 \; ,
 \label{eq:Euler}
 \end{equation}
where $\sigma_{\pm} =  \frac{1}{2 } [ \sigma_{1} \pm i \sigma_{2} ]$, and
the three functions
$\Phi_{\pm} ( x )$, 
$\Phi_{3} ( x )$ 
have to be chosen such that $U ( x )$ satisfies
Eq.(\ref{eq:Udif}). 
A parameterization  similar to Eq.(\ref{eq:Euler}) has
recently been used by
Schopohl\cite{Schopohl98} 
to study the Eilenberger equations 
of superconductivity.
We find that
the ansatz (\ref{eq:Euler}) solves Eq.(\ref{eq:Udif})
if $\Phi_{\pm} ( x )$ and $\Phi_{3} ( x )$
satisfy
\begin{mathletters}
 \begin{eqnarray}
 \partial_x \Phi_{+} & = & - 2 i \omega \Phi_{+} + \Delta ( x ) 
 [ 1 - \Phi_{+}^2 ] 
 \label{eq:phiplus}
 \; ,
 \\
 \partial_x \Phi_{-} & = &  2 i \omega \Phi_{-} - \Delta ( x ) 
 [ 1 - 2 \Phi_{+} \Phi_{-} ] 
 \label{eq:phiminus}
 \; ,
 \\
  \partial_x \Phi_{3} & = & -i \Delta ( x ) \Phi_{+}
 \label{eq:phi3}
 \; .
 \end{eqnarray}
\end{mathletters}
Non-linear differential equations
of the type (\ref{eq:phiplus}) are called Riccati equations.
The set of equations 
obtained by Schopohl\cite{Schopohl98} 
has a similar structure 
but is not identical with Eqs.(\ref{eq:phiplus}-\ref{eq:phi3}).
Note that Eq.(\ref{eq:phiplus}) involves only $\Phi_{+}$.
If we manage to obtain the solution $\Phi_+$,
Eqs.(\ref{eq:phiminus},\ref{eq:phi3}) 
become simple linear equations which can be solved exactly.
From Eqs.(\ref{eq:localdos},\ref{eq:ansatz}) and (\ref{eq:Euler})
it is easy to see that the local DOS can be written as
 \begin{equation}
 \rho ( x , \omega ) = {\pi}^{-1} {\rm Re} R ( x , \omega + i 0 )
 \; \; , \; \; 
 R  = 1 - 2 \Phi_{+} \Phi_{-} 
 \; .
 \label{eq:rhoR}
 \end{equation}
Thus, 
to calculate the  average DOS
we have to average
the product $\Phi_{+} \Phi_{-}$ 
over the probability distribution of the
field $\Delta ( x )$. 
In the limit  where the right-hand-side of Eq.(\ref{eq:cov})
reduces to
$2 D \delta ( x - x^{\prime})$ 
we can use the fact that
$\Phi_{+}$ and $\Phi_{-}$
satisfy first order differential equations 
to express the average $\langle \Phi_{+} \Phi_{-} \rangle$
in terms of the solution of two coupled one-dimensional
Fokker-Planck equations\cite{Bartosch98}.
Thus, our method leads to an algorithm
for obtaining the exact
$\langle \rho ( \omega ) \rangle$ 
without using the node counting 
theorem\cite{Lifshits88}.
However, the
Fokker-Planck equation obtained by
OE\cite{Ovchinnikov77} within the phase formalism\cite{Lifshits88}
is easier to solve than our system of
two coupled Fokker-Planck equations.
Hence, 
for $\delta$-function correlated disorder
our approach does not have any practical advantage.

On the other hand, if the
disorder is not $\delta$-function
correlated, probability distributions 
of physical quantities
do in general not satisfy 
Fokker-Planck equations, and
it is not so easy to 
perform controlled calculations
or even obtain exact results\cite{vanKampen81}.
We now show that for
real $\Delta ( x)$
the local DOS at the Fermi energy 
can be calculated exactly.  
To derive this result, 
let us introduce the
complex vector
 \begin{equation}
  \vec{\psi} =
  \left(
 \begin{array}{c}
  - \sqrt{2}  ( 1 - \Phi_{+} \Phi_{-} ) \Phi_{+}
  \\
  1 - 2 \Phi_{+} \Phi_{-}
  \\
  \sqrt{2}  \Phi_{-}
 \end{array}
 \right) 
 \equiv
 \left(
 \begin{array}{c}
 Z_{+}  \\ R \\ Z_{-} 
 \end{array}
 \right) 
 \; .
 \label{eq:psidef}
 \end{equation}
Note that by construction
$R^2 -  2 Z_{+} Z_{-} = 1 $ for all $x$, and that 
according to Eq.(\ref{eq:rhoR})
the second component of $\vec{\psi}$ is related to the local
DOS. Using
Eqs.(\ref{eq:phiplus},\ref{eq:phiminus}) we find 
 \begin{equation}
  \partial_x \vec{\psi} = - H ( x )  \vec{\psi}
  \; \; , \; \; 
 H (x ) =  2 i \omega J_3 + 2 \Delta ( x ) J_1
 \; , 
 \label{eq:multistoch}
 \end{equation}
where $J_i$ are spin $J=1$ operators in the representation
 \begin{equation}
 J_3 =
 \left(
 \begin{array}{ccc}
   1   & 0 & 0 \\
  0 &  0 &  0 \\
   0 &  0 &  -1 
  \end{array}
  \right)
  \; \;  , \;  \; 
 J_1 = \frac{1}{\sqrt{2}}
 \left(
 \begin{array}{ccc}
   0   & 1 & 0 \\
  1 &  0 &  1 \\
   0 &  1 &  0 
  \end{array}
  \right)
  \; . 
  \label{eq:Ldef}
  \end{equation}
Eq.(\ref{eq:multistoch}) is a
linear multiplicative stochastic 
differential equation\cite{vanKampen81}.
Formally this equation looks like
the imaginary time  Schr\"{o}dinger equation
for a $J = 1$ quantum spin 
in a random magnetic field,
with 
$x$ playing the role of imaginary time.
Although the operator $H$ in Eq.(\ref{eq:multistoch}) is
not hermitian, we may 
perform an analytic continuation to imaginary
frequencies ($\omega = i E$) to obtain a
hermitian spin Hamiltonian.
We thus arrive at the remarkable result
that the average DOS of the FGM  
can be obtained  from the
{\it{average state-vector}} of a  $J = 1$ 
spin in a random magnetic field.
Our non-linear transformation
(\ref{eq:psidef})
is well known
in the quantum theory of magnetism:
with the formal identification $  R \rightarrow J_3$, $ \sqrt{2}  Z_{\pm} 
\rightarrow \mp J_{\pm}$,
$\sqrt{2} \Phi_{-} \rightarrow b^{\dagger}$, 
and $\sqrt{2} \Phi_{+} \rightarrow b$,
Eq.(\ref{eq:psidef}) is precisely the
Dyson-Maleev transformation\cite{Dyson56}, which 
expresses
the spin operators in terms of boson operators $b$,$b^{\dagger}$.

The solution of Eq.(\ref{eq:multistoch}) with initial
condition $\vec{\psi} ( 0 ) = \vec{\psi}_0$ 
is
 $\vec{\psi} ( x ) =  S ( x ) \vec{\psi}_0$,
where the $S$-matrix is
 \begin{equation}
 S ( x ) = {\cal{T}} \exp \left[ -  { 
 \int_{0}^{x} d x^{\prime} H ( x^{\prime} ) } 
 \right]
 \; .
 \label{eq:timeord}
 \end{equation}
Here ${\cal{T}}$ is the usual time ordering 
operator.
The proper choice of 
boundary conditions requires some care.
For simplicity, let us assume that $\Delta ( x )$ 
is non-zero only
in a finite interval $0 \leq x \leq L$.
Outside this regime we find that
 $Z_{\pm} ( x ) = \exp [ \mp 2 i ( \omega + i 0) (x - x_0)  
 ]
 Z_{\pm} ( x_0 )$ is the solution of Eq.(\ref{eq:multistoch}).
%(implying $\rho ( x, \omega ) = 1 / \pi$ for
%$x < 0$ and $x > L$)
Physically it is clear that exponentially
growing solutions are forbidden, which requires
$Z_{-} ( 0 ) = Z_{+} ( L ) = 0 $ and implies $R = 1$ for
$x \leq 0$ and $x \geq L$.
We conclude that at the boundaries 
 \begin{equation}
  \vec{\psi}_0 =
 \left(
 \begin{array}{c}
 Z_{+}(0)  \\ 1 \\ 0 
 \end{array}
 \right) 
 \; \; ,
 \; \; 
  \vec{\psi} ( L ) =
 \left(
 \begin{array}{c}
 0  \\ 1 \\ Z_{-} ( L)  
 \end{array}
 \right) 
 \; ,
 \label{eq:boundary}
 \end{equation}
where $Z_{+} ( 0 )$ and $Z_{-} ( L)$
are determined by 
$\vec{\psi} ( L ) = S ( L ) \vec{\psi}_0$.
This implies $Z_{+} ( 0 ) = -  S_{12} ( L ) / S_{11} ( L )$.
Because the matrix elements of $S ( L )$ depend on the disorder,
the initial vector $\vec{\psi}_0$ is stochastic. 
Note that in textbook discussions
of multiplicative stochastic differential equations
one often assumes deterministic initial conditions\cite{vanKampen81}.
%Thus, 
%the average $\langle \vec{\psi} ( x ) \rangle$
%cannot be expressed in terms of the average 
%propagator $\langle G ( x ) \rangle$.
After simple algebra we obtain for the
second component of $\vec{\psi} ( x ) = S ( x ) \vec{\psi}_0$ 
in the interval $0 \leq x \leq L$
 \begin{equation}
 R ( x ) = S_{22} ( x ) - S_{21} ( x ) S_{12} ( L )/S_{11} ( L )
 \; .
 \label{eq:Rres}
 \end{equation}
In general we have to rely on approximations
to calculate the time-ordered exponential in Eq.(\ref{eq:timeord}).
However, there are
two special cases where $S ( x )$ can be calculated exactly.
The first is obvious: if $\Delta ( x ) = \Delta_0 $ is independent
of $x$, our spin Hamiltonian $H (x )$ is 
constant, so that the time-ordering
operator is not necessary. 
Eq.(\ref{eq:Rres}) can then be 
evaluated exactly for arbitrary $L$\cite{Bartosch98}. 
If we take the limit $L \rightarrow \infty$ holding
$x / L$ fixed,
we recover the well known square root singularity
at the band edges, 
 \begin{equation}
 \lim_{L \rightarrow \infty} \rho ( x   , \omega ) = 
 \frac{ \Theta ( \omega^2 - \Delta_0^2 )
  | \omega | }{\pi \sqrt{ \omega^2 - \Delta_0^2 }}
  \;  \; , \; \; 0 < x/L < 1
  \; .
 \label{eq:deltaconst}
 \end{equation}
There exists another, more interesting 
limit
where $S ( x )$ can be calculated exactly.
Obviously, at $\omega = 0$ the {\it{direction}}
of the magnetic field in our 
spin Hamiltonian (\ref{eq:multistoch}) is constant.
Although in this case $H ( x )$  is $x$-dependent,
we may omit the time-ordering operator
in Eq.(\ref{eq:timeord}). After straightforward algebra
we obtain
from Eq.(\ref{eq:Rres})
 \begin{equation}
  R ( x )  =
 {\cosh [ A ( x ) - B ( x ) ] }/{
 \cosh [ A ( x ) + B  ( x ) ] }
 \; ,
 \label{eq:Rdis}
 \end{equation}
where
$ A ( x ) = \int_{0}^{x} d x^{\prime} \Delta ( x^{\prime} )$
and
$B ( x ) = \int_{x}^{L} d x^{\prime} \Delta ( x^{\prime} )$.
In Eq.(\ref{eq:Rdis}) it is
understood that
$R ( x )$ stands for $R ( x ,  i 0 )$, so that
$\rho ( x , 0 ) = \pi^{-1} R (  x )$.
We have thus succeeded to calculate 
the local DOS $\rho ( x , \omega = 0 )$ of the FGM
at the Fermi energy {\it{for a given
realization of the disorder.}}
The special  symmetries 
of random Dirac fermions at $\omega = 0$
have recently been used by
Shelton and Tsvelik\cite{Shelton97} to
calculate the statistics of the corresponding
wave-functions.

To calculate 
the disorder average of Eq.(\ref{eq:Rdis}),
we introduce 
%the joint distribution of
%$A ( x )$ and $B  (x)$, 
% \begin{equation}
$  P ( x ; a , b )  
  = \langle \delta ( a - A ( x ) ) \delta ( b - B ( x ) )
  \rangle
  $
%  \label{eq:joint}
%  \; ,
%  \end{equation}
and write
 \begin{equation}
 \langle R ( x ) \rangle 
 = \int_{- \infty}^{\infty} d a
 \int_{- \infty}^{\infty}
  d b
  P ( x ; a , b ) \frac{ 
  \cosh ( a - b  ) }{ \cosh   ( a+ b ) }
  \; .
  \label{eq:Rav}
  \end{equation}
Assuming that the probability distribution of $\Delta ( x )$ is Gaussian
with average $\Delta_0$ and covariance $K ( x , x^{\prime} )$,
the joint distribution $P ( x ; a, b )$ can be calculated exactly.
The integration over the difference $a - b$ in Eq.(\ref{eq:Rav}) can 
then be performed, and we obtain
for the average local DOS $\langle \rho ( x , 0 ) \rangle = \pi^{-1}
\langle R ( x ) \rangle$,
 \begin{eqnarray}
 \langle \rho ( x , 0 ) \rangle
 & = & \frac{ e^{  \alpha (x )} }{ \pi \sqrt{ 2 \pi \sigma^2}}
 \int_{- \infty}^{\infty} ds \exp \left[  - s^2 / (2 \sigma^2 ) \right]
 \nonumber
 \\
 & \times &
 \frac{ \cosh [ \beta ( x ) s + \Delta_0 ( 2x - L ) ]}{
 \cosh [ s + \Delta_0 L ] }
 \; .
 \label{eq:Rav1}
 \end{eqnarray}
Here 
%$\sigma^2 = C_1  + C_2 + 2 C_3$,
$\sigma^2 = C_1 (L)$,
$\alpha ( x ) = 2 [ C_1(x) C_2(x) - C_3^2(x)]/ \sigma^2$, and
$\beta ( x ) = [ C_1(x) - C_2(x) ] / \sigma^2$, with
 \begin{equation}
 C_1 ( x ) = \int_{0}^{x} d x^{\prime} 
 \int_{0}^{x} d x^{\prime \prime }  K ( x^{\prime} , x^{\prime \prime } )
 \; .
 \label{eq:C1def}
 \end{equation}
$C_{2} ( x )$ is defined by replacing
the range of the integrals in Eq.(\ref{eq:C1def}) 
by the interval $[x , L]$, and $C_3 ( x )$
is obtained by  choosing
the interval
$ [ 0 , x ]$ for the $x^{\prime}$- and
$[x, L]$ for the $x^{\prime \prime}$-integration. 
%By construction $\sigma^2$ is independent of $x$.

\begin{figure}
\begin{center}
%\vspace{1cm}
%\epsfysize4.0cm 
%\hspace{3mm}
%\epsfbox{fig1.eps}
%\vspace{3mm}
%\psfrag{xxx}{$x/L$}
%\psfrag{RRRRRR}{$\langle R ( x , i 0 ) \rangle$}
 \epsfbox{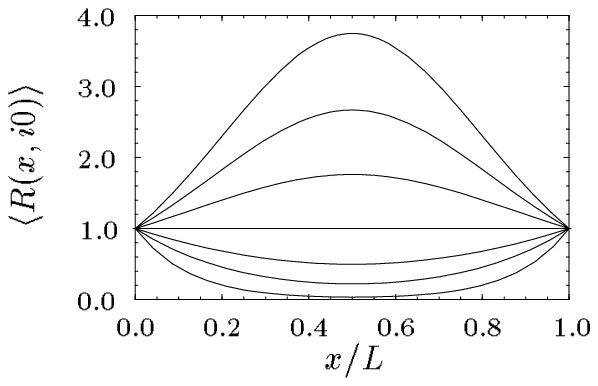} %,height=4.2cm}
\vspace{2mm}
\caption{Disorder average $ \langle R ( x , i 0 ) \rangle
= \pi \langle \rho ( x , 0 ) \rangle $ for
$\tilde{L} = 4$ and $\lambda =0$ as
function of $x / L$, see Eq.(\ref{eq:Rav1}).
From top to bottom: $\nu = 0, 0.5, 0.75,
1, 1.25, 1.5, 2$.}
\label{fig:Rx}
\end{center}
\end{figure}

\begin{figure}
\begin{center}
%\vspace{1cm}
%\epsfysize4.0cm 
%\hspace{3mm}
%\psfrag{f}{$f(\lambda)$}
%\psfrag{nu}{$\nu_{c}(\lambda)$}
%\psfrag{x}{$\lambda$}
%\psfrag{ffffnnnn}{$f(\lambda)$,\ $\nu_{c}(\lambda)$}
\epsfbox{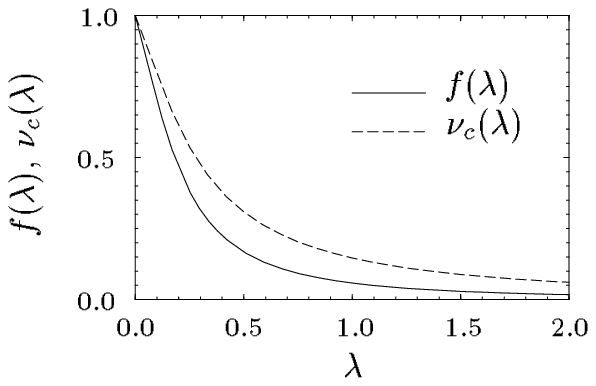} %,height=4.2cm}
\vspace{2mm}
\caption{
Plot of $f ( \lambda )$ 
and $\nu_c ( \lambda )$ defined in
Eqs.(\ref{eq:fdef}) and (\ref{eq:nuc}).
%Dashed line: the critical value $\nu_c ( \lambda )$ defined in
%Eq.(\ref{eq:nuc}). For $\lambda \rightarrow \infty$ we find 
%Solid line:
%the function $f ( \lambda )$ defined in Eq.(\ref{eq:fdef}).
%Dashed line: the critical value $\nu_c ( \lambda )$ defined in
%Eq.(\ref{eq:nuc}). For $\lambda \rightarrow \infty$ we find 
%$f ( \lambda )  \sim 1 / (12  \lambda^{2})$ and
%$\nu_c ( \lambda ) \sim 
% ( 24 \lambda^{3} )^{-1/2}$. 
}
\label{fig:fplot}
\end{center}
\end{figure}

We now specify 
$K ( x ,  x^{\prime})$ to be of the form (\ref{eq:cov}).
Then
$\sigma^2$, $\alpha ( x )$, and $\beta (x )$
are easily calculated.
It is convenient to introduce the dimensionless
parameters $\tilde{L} = 2 K_0 \xi L$, $\lambda = \xi / L$, and
$ \nu = \Delta_0 / (2 K_0 \xi )$.
Numerical results for $\langle \rho ( x , 0 ) \rangle$ 
%for representative values of these parameters 
are shown in
Fig.\ref{fig:Rx}.
Due to symmetry with respect to $x = L / 2$ the local DOS 
assumes an extremum at $x = L / 2$, which in the limit $L \rightarrow \infty$
approaches either zero or infinity.
Using the fact that at $x = L/2$ the $\cosh$ in the numerator
of Eq.(\ref{eq:Rav1}) is unity,  we obtain
 \begin{equation}
 \langle \rho  ( {L}/{2} , 0 ) \rangle = 
 \frac{ e^{ \frac{\tilde{L}}{2} f ( \lambda ) }}{ \pi \sqrt{2 \pi \sigma^2 }}
 \int_{- \infty}^{\infty} ds \frac{ \exp \left[ { - s^2 / (2 \sigma^2 )} \right]}{
 \cosh [ s + \nu \tilde{L}  ] }
 \; ,
 \label{eq:Rhalf}
 \end{equation}
where  $\sigma^2 = \tilde{L} [ 1 - \lambda ( 1 - e^{- 1 / \lambda} ) ]$
and
 \begin{equation}
 f ( \lambda ) = 
 1 - \lambda
 [ 3 - 4 e^{- 1 / (2 \lambda )} + e^{- 1 / \lambda} ] 
 \label{eq:fdef}
 \; .
 \end{equation}
As shown in Fig.\ref{fig:fplot}, 
$f ( \lambda )$ is positive
and  monotonically decreasing. 
%with $f ( 0) = 1$.
Let us first consider the case $\nu = 0$.
This corresponds to
the model discussed by Sadovskii\cite{Sadovskii79}
with real $\Delta (  x)$.
For $\tilde{L} \gg 1$ the $s$-integration in Eq.(\ref{eq:Rhalf}) 
is easily done, and we find 
 $\langle \rho  ( \frac{L}{2} , 0 ) \rangle  \propto 
 \tilde{L}^{-1/2} \exp[ \frac{\tilde{L}}{2} f ( \lambda )  ]$.
Keeping in mind that for any finite
$\xi$  the  parameter $\lambda = \xi / L$ 
vanishes for $L \rightarrow \infty$, it is
obvious that in this limit 
$\langle \rho ( L/2 , 0 ) \rangle$ is infinite.
From Eq.(\ref{eq:Rav1}) it is easy
to show numerically 
that this is also true for 
$\lim_{L \rightarrow \infty} \langle \rho ( x , 0 ) \rangle$ 
in the open interval $0<x/L<1$.
We have thus proven that for $L \rightarrow \infty$
and arbitrary $\xi < \infty$
the average DOS of the FGM is infinite at the Fermi energy,
in agreement
with general symmetry arguments\cite{Lifshits88,Mostovoy98}.
%with general symmetry
%Because for large $\xi < \infty$ the algorithm
%constructed by Sadovskii\cite{Sadovskii79}
%predicts a pseudogap in the DOS  
%instead of a singularity, we are lead to the
%conclusion that,
%at least for real $\Delta ( x )$, 
%this algorithm is not reliable.

For finite $\nu$
a careful analysis\cite{Bartosch98} of Eq.(\ref{eq:Rhalf}) shows that
there exists a critical value $\nu_c ( \lambda )$ such
that for $| \nu | > \nu_c$ 
the local DOS
$\langle \rho ( L/2 , 0 ) \rangle $ scales to
zero in the thermodynamic limit, and a pseudogap
emerges. 
We obtain
 \begin{equation}
 \nu_c ( \lambda )    =  [ 1 - \lambda ( 1 - e^{- 1/ \lambda} ) 
 ]^{1/2} [ f ( \lambda ) ]^{1/2}
 \; ,
 \label{eq:nuc}
 \end{equation}
%A graph of $\nu_c ( \lambda )$ is shown in Fig.\ref{fig:fplot}.
see Fig.\ref{fig:fplot}.
For $\lambda = 0$
Eq.(\ref{eq:nuc}) yields $\nu_c ( 0 ) = 1$.
At the first sight this seems to contradict
the result of
OE\cite{Ovchinnikov77}, who found pseudogap behavior already
for $|\nu | > 1/2$.
%\cite{OEfoot}.
One should keep in mind, however, that we have
set $\omega = 0$ {\it{before}} taking the limit $L \rightarrow \infty$,
while in Ref.\cite{Ovchinnikov77}
these limits are taken in the opposite order.
The non-commutativity of these limits 
is well known from
the calculation of the local DOS 
of the Tomonaga-Luttinger model
with a boundary\cite{Eggert96}. 
Interestingly, 
for $K ( x , x^{\prime} ) = 2 D \delta ( x - x^{\prime})$
there exists a regime 
$1/2 < \nu  < 1$ where
in the thermodynamic limit
$\langle \rho ( \omega ) \rangle$ is
discontinuous at $\omega = 0$. 
This follows from the fact that
according to OE 
$\langle \rho ( \omega ) \rangle \propto | \omega |^{2 \nu -1}
\rightarrow 0$ for $\omega \rightarrow 0$, 
whereas we have shown that $\langle \rho ( 0 ) \rangle = \infty$.
For $\nu = 0$ and large but finite $\xi$ we 
conjecture
that the frequency-dependence
of $\lim_{L \rightarrow \infty} \langle 
\rho ( \omega ) \rangle$  
is qualitatively similar to the behavior found
by Fabrizio and M\'{e}lin\cite{Fabrizio97}
for random Dirac fermions 
with a special type of disorder\cite{Comtet95}.
Specifically, we expect that 
for frequencies exceeding a certain 
crossover frequency $\omega^{\ast}$
the average DOS of the FGM shows pseudogap behavior,
which is correctly predicted by
Sadovskii's algorithm\cite{Sadovskii79}. 
However, this algorithm misses the Dyson singularity,
which emerges 
for frequencies
$| \omega | 
{ \raisebox{-0.5ex}{$\; \stackrel{<}{\sim} \;$}}  \omega^{\ast}$ 
for any finite value of $\xi$.

In summary, we have developed a non-perturbative method
to calculate the Green's function of the FGM.  
Our main result is the proof of the existence
of a singularity
in $\langle \rho ( 0 ) \rangle$ 
for any finite value of the correlation
length $\xi$ as long as $\Delta (x)$ is real and
$| \Delta_0 |$ is sufficiently small.
For $\Delta_0 =0$ we have shown that
with open boundary conditions
$\langle \rho ( \frac{L}{2}, 0 ) \rangle
\propto \exp [ K_0 \xi L f ( \xi / L ) ]$,
where $f ( 0 ) = 1$. 
Moreover, if we let $L \rightarrow \infty$ keeping 
$x/L \in ( 0,1)$  fixed,
$\langle \rho ( x , 0 ) \rangle$ 
exhibits a similar singularity\cite{Bartosch98}.
%Because for finite $\omega$ this 
%singularity persists at least up to frequencies
%of order $\pi / L$, 
%the integral $\int d \omega
%\langle \rho ( \omega ) \rangle$ is 
%for $L \rightarrow \infty$ 
%{\it{infrared}} 
%divergent. 
Thus, disorder pushes
states from high energies to the band center.
For finite $\Delta_0$ this effect competes with
the suppression of the DOS due to long-range order.
In the incommensurate case (where $\Delta(x)$ is complex and in
Eq. (\ref{eq:matrix}) 
we should replace  $i \sigma_2 \Delta \rightarrow \sigma_{+} \Delta - 
\sigma_{-} \Delta^{\ast}$ ) it is known that 
$\langle \rho (0) \rangle$ is finite
in the white-noise limit\cite{McKenzie96,Fischbeck90}.   
We expect that this remains true for
arbitrary $\xi$.  
In fact, there exists numerical evidence that
in the incommensurate case $\langle \rho ( \omega ) \rangle$
can be accurately calculated from 
Sadovskii's algorithm\cite{Monien98}.
For a comparison with experiments
one should keep in mind that 
any violation of the perfect charge conjugation symmetry
%(for example due to
%forward scattering by impurities\cite{Ovchinnikov77})
will wash out the singularity at $\omega = 0$.
It is therefore unlikely that the singularity
is visible in realistic materials, although an enhancement
might survive.
%in systems where 
%the charge conjugation symmetry is only
%weakly broken.
The fact that 
the singular behavior of
$\langle \rho (0 ) \rangle$
in the FGM with real $\Delta (x)$
is only destroyed if
$|\nu| = |\Delta_0| / ( 2 K_0 \xi )$
exceeds a finite critical value implies
that in
{\it{commensurate}} Peierls chains 
true
pseudogap behavior should emerge gradually
below the Peierls transition
when the order parameter $\Delta_0$
is sufficiently large.
Lee, Rice, and Anderson\cite{Lee73} 
came to a similar conclusion within 
perturbation theory.

We thank M. V. Sadovskii,
K. Sch\"{o}nhammer, R. Hayn, R. H. McKenzie, and H. Monien for discussions and
comments.
%We also thank R. H. McKenzie and
%M. V. Sadovskii for pointing out that in the incommensurate case
%$\langle \rho (0) \rangle$ is expected to remain finite.
This work was financially supported by the
DFG (Grants No. Ko 1442/3-1 and Ko 1442/4-1).

%\newpage
%\widetext
% 

%
%
%
\vspace{-5mm}

\begin{thebibliography}{99}
\bibitem{footnote1}
We measure energies relative to the Fermi energy
and set the Fermi velocity and $\hbar$ equal to unity.
\end{thebibliography}
\end{document}